\documentclass[aip, 
reprint, 
twocolumn
]{revtex4-1}
\usepackage[english]{babel}
\usepackage{amsmath}
\usepackage{amsfonts}
\usepackage{amssymb}
\usepackage{graphicx}
\usepackage{subfigure}

\begin{document}
\author{Tommaso Lunghi}
\email[]{Tommaso.Lunghi@unige.ch}
\author{Enrico Pomarico}
\author{Claudio Barreiro}
\affiliation{Group of Applied Physics, University of Geneva, Geneva CH-1211, Switzerland}
\author{Damien Stucki}
\affiliation{ID Quantique SA, Rue de la Marbrerie 3, CH-1227 Carouge, Switzerland}
\author{Bruno Sanguinetti}
\author{Hugo Zbinden}
\affiliation{Group of Applied Physics, University of Geneva, Geneva CH-1211, Switzerland}
\title{Advantages of gated silicon single photon detectors}

\begin{abstract}
We present a gated silicon single photon detector based on a commercially available avalanche photodiode. Our detector achieves a photon detection efficiency of 45$\pm$5\% at 808\,nm with 2$\cdot$10$^{-6}$ dark count per ns at -30V of excess bias and -30$^{\circ}$C. We compare gated and free-running detectors and show that this mode of operation has significant advantages in two representative experimental scenarios: detecting a single photon either hidden in faint continuous light or after a strong pulse. We also explore, at different temperatures and incident light intensities, the ``charge persistence'' effect, whereby a detector clicks some time after having been illuminated.
\end{abstract}
\maketitle
\section{Introduction}
Silicon single-photon avalanche diodes (Si SPADs) are a standard solid-state solution for single-photon detection in the visible and near-infrared\cite{CovaJMO51}. 
In particular, Si SPADs can attain high photon-detection efficiencies and extremely low dark-count rates. 
The structure based entirely on silicon has only a limited number of traps in the multiplication region, resulting in a device that is less affected by afterpulsing compared to the InGaAs/InP detectors in the telecom range. Thus, Si SPADs are normally used in free-running mode.
However, recently the advantages of gating a thin Si diode has been explored for near-infrared spectroscopy experiments\cite{DallaMoraSTQE16}.

\begin{figure}[!tbp]
	\includegraphics[width=7 cm]{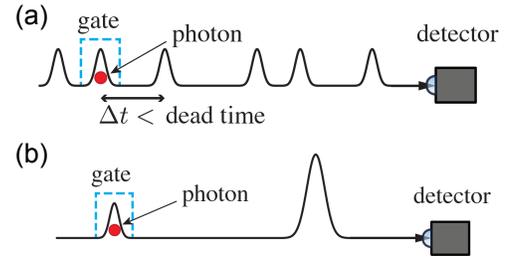}
	\caption{The experimental scenarios where a gated detector plays an essential role.(a): applying a gate ensures that the detector is active when the photon of interest arrives, even if the detector is constantly being illuminated. (b): applying a gate ensures that the detector is not blinded by the preceding strong pulse.}
	\label{fig:1}
\end{figure}

Here we identify two general situations where a gated Si detector shows some essential advantages:\\
\emph{(A) Detecting a photon hidden in continuous faint light}: in this scenario (Fig.\ref{fig:1}(a)) the detector is continuously illuminated by faint light. The time interval between two subsequent photons is, on average, smaller than the dead time of the detector. The photons of interest are hidden in the beam but their arrival times are known.
Under these conditions, a free-running detector is saturated or even blinded, reducing the detection probability. This situation can be found in quantum-cloning\cite{Lamas-LinaresS296} and faithful swapping\cite{SangouardPRL106} experiments. \\
\emph{(B) Detecting a photon arriving after a strong pulse}: in this scenario (Fig.\ref{fig:1}(b)) a strong pulse impinges on the detector before the arrival of the photon of interest. Hence, when the photon of interest arrives, a free-running detector is either in its dead time or its noise is highly increased by afterpulsing. This situation can be found in Optical Time Domain Reflectometry\cite{EraerdsJLT28}, in fluorescence spectroscopy\cite{DallaMoraSTQE16} as well as in quantum-memory experiments which require strong preparation pulses \cite{TimoneyJPB}.

In this paper we describe the performance of our gated Si SPAD for different temperatures and light intensities. Finally, we show the advantages of using a gated detector with respect to free-running ones in the scenarios \textbf{A} and \textbf{B}.\\

\section{The Gated Si APD module}
In our work we use a commercially available Excelitas C30921SH Si APD. This diode is based on a {\it reach-through} structure \cite{WebbRCA35}, and is characterized by a large detection area (0.5\,mm diameter) and high detection efficiency between 600 and 1050\,nm with a maximum at 800\,nm. Previous work \cite{kimRSI82} has detailed its performance in terms of detection efficiency and dark-count rate when passively quenched. The high efficiency comes at the cost of a high excess bias ($\sim$40 V) with a breakdown voltage of the order of 230 V at room temperature. Better performances could be achieved with smaller diodes, however for commercial reasons, these are only sold in modules.
 
\begin{figure}[!htbp]
	\includegraphics[width=7cm]{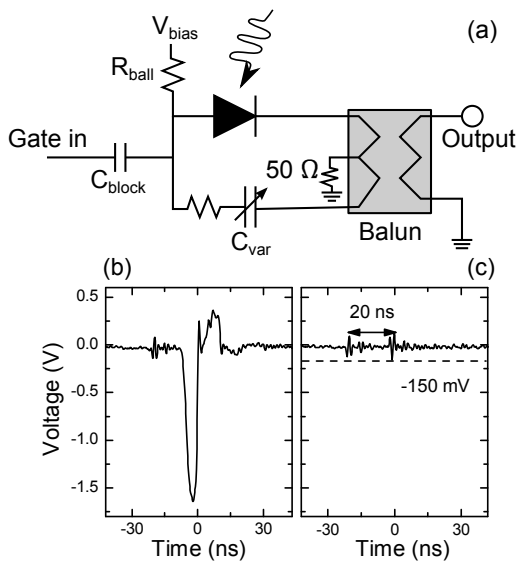}
	\caption{(a) Electrical circuit. (b) Avalanche measured with a 500 MHz oscilloscope. (c) No avalanche, the derivative peaks at -20 ns and 0 ns have $\leq$150\,mV of amplitude.}
	\label{fig:2}
\end{figure}

The electrical circuit that drives the diode is outlined in Fig.\ref{fig:2}(a). The detector is initially biased with a (negative) voltage, V$_{bias}$, below the breakdown voltage, V$_{bd}$. Single photon counting is enabled by applying a square gate voltage V$_{gate}$ on the detector anode. By choosing a suitable value for the blocking capacitor C$_{block}$ (1\,nF), the anode voltage is kept constant for the duration of the gate. An Avtech pulser (AVI-MP-N) produces a clean gate with variable duration at a repetition rate of up to 1 MHz. We employ repetition rates of 10 KHz and 100 KHz and gate amplitudes of 37.6 V and 36.4 V respectively. The rise and fall times of the gate are below 1\,ns. In our characterization we used 20\,ns gates, although we tested gates down to 5\,ns. Due to the intrinsic capacitance of the diode a derivative signal is generated at the cathode consisting of two spikes of 7 V in correspondence of the rising and the falling edges of the gate. An auxiliary line, simulating the capacitance and resistance of the detector is introduced to produce equal spikes which are then subtracted from the signal line using a balun (Mini Circuits RF transformer, TCM4-1W). The avalanche, which has an amplitude larger than 1 V is easily discriminated from the residual spikes which, after the balun, have an amplitude smaller than 150\,mV. Figures \ref{fig:2}(b) and \ref{fig:2}(c)  show the output signal with and without the avalanche.%
\section{Results}
The detection efficiency, $\eta$, and the dark count probability per gate, $p_{dc}$, are measured using a pulsed laser diode at 808 nm with a repetition rate of 100 KHz. The laser beam has a spot size of 25 $\mu$m of diameter on the free space detector and the intensity is attenuated to on average of 1 photon per pulse.
$\eta$ is evaluated taking into account the Poissonian photon-number distribution of the pulse\cite{thewAPL91}.

\begin{figure}[!htbp]
	\centering
	\includegraphics[width=8 cm]{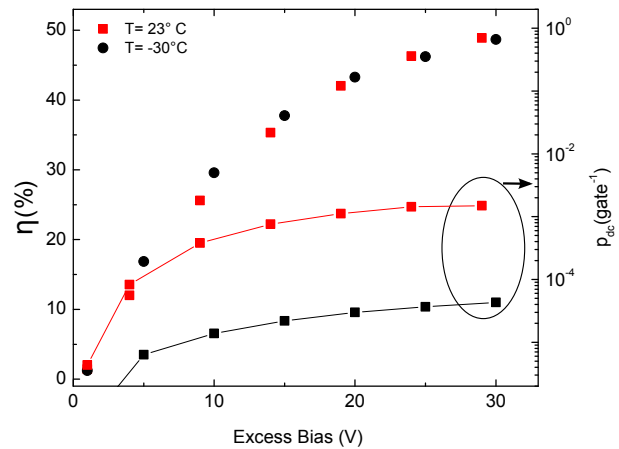}
	\caption{Detection efficiency and dark count probability per gate (20 ns) as a function of the excess bias V$_{exc}$ = V$_{bias}$ - V$_{bd}$ at temperatures of 23$^{\circ}$C and -30$^{\circ}$C.
	}
	\label{fig:3}
\end{figure}

Figure \ref{fig:3} shows $\eta$ and p$_{dc}$ as a function of excess bias V$_{exc}$ = V$_{bias}$ - V$_{bd}$, for temperatures of both -30$^{\circ}$C and 23$^{\circ}$C.
The detection efficiency at 808\,nm is of 45\%$\pm$5\% with dark counts per gate of 4$\cdot 10^{-5}$ at 30 V of excess bias and -30$^{\circ}$C.\\

To evaluate the detector performance in experiments that fall within the scenario {\bf (B)}, we investigate the effects that strong pulses produce on the dark-count probability. This effect generates noise in experiment that fall within the scenario (B) described in the introduction. 
 
\begin{figure}[!htbp]	
	\includegraphics[width=6.5cm]{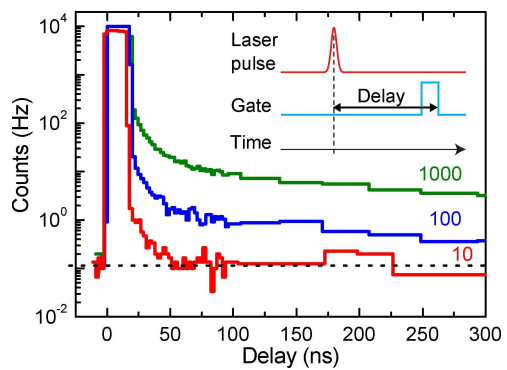}
	\caption{Count rate as a function of the delay between the end of the rising edge of the gate and the laser pulse (see inset) for different pulse energies (wavelength=655 nm).
	The black dotted line indicates the dark count rate. 
	}
	\label{fig:4}
\end{figure}

A programmable delay generator (SRS DG535) triggers a 655 nm laser (55 ps FWHM, Picoquant LDH-P-650) and the detector. The laser light is attenuated down to an average of 10/100/1000 photons per pulse. The number of detections (avalanches occurring during these gates) are recorded versus the delay between the falling edge of the gate and the laser pulse (see the inset of Fig. \ref{fig:4}): a zero delay corresponds to the synchronization between the falling edge of the gate pulse and the arrival of the laser light on the detector.

\begin{figure}[!htbp]
	\centering
	\includegraphics[width=6.5 cm]{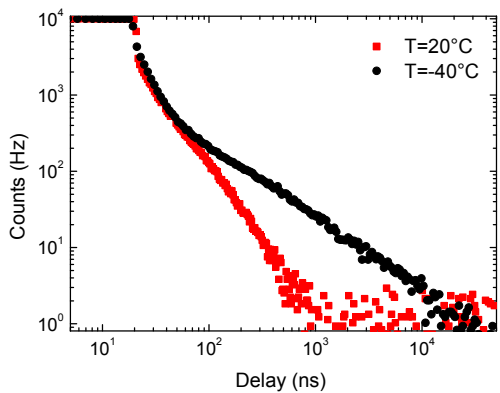}
	\caption{Time-response for two different temperatures, i.e. 20$^\circ$C and -40$^\circ$C after subtraction of the dark count rate (wavelength=655 nm). }
	\label{fig:5}
\end{figure}

The results are presented in Fig.\ref{fig:5}: for negative delays, the gate ends before the arrival of the laser pulse and the count rate drops to the intrinsic dark count rate. If the laser pulse impinges on the detector during the gate the number of detections saturates to the repetition rate of 10 KHz. For longer delays, the gate arrives after the laser but the light can still provoke an avalanche when the gate is applied. 
We verified that this noise is proportional to the energy of the laser pulse.

What is the origin of this noise? It is not due to afterpulsing since no avalanche is generated when the photons hit the detector. Neither we can relate it to slow diffusion of photoelectrons in the diode: the absorption region in {\it reach-through} structure is filled by an electric field which forces every photo-generated electron towards the multiplication region. This is confirmed by time-jitter measurements, which show a FWHM of 500\,ps and, in particular, no long tail\cite{SpinelliED44} (contrary to thin diodes\cite{RipamontiSSE28}). However, the defects in all the absorption region may trap photo-generated electrons as occurs in the multiplication region for afterpulsing. The release of these electrons can happen during the gate, thus giving rise to an avalanche. This noise source is called charge persistence\cite{ZhangJQO45}. 

We investigate the charge persistence effect for two different temperatures with 10$^5$ photons per pulse (see Fig.\ref{fig:5}). The result appears to confirm the explanation provided: since the lifetime of the electrons trapped increases at lower temperature, the decay time is longer.

\section{Discussion}\label{Sec:IV}
In this section we compare gated and free-running operations of SPADs in the two scenarios.\\ 

\begin{figure}[!htbp]
	\centering
	\includegraphics[width=7 cm]{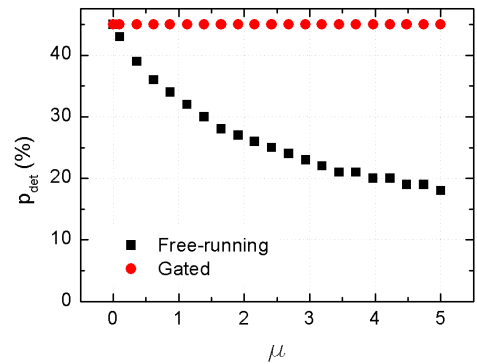}
	\caption{{\bf Scenario (A):} detection probability vs average number of photon per $\Delta$t, for the different operational modes. E.g. for $\Delta$t=32 ns, $\mu$=5 corresponds to 47 pW (156$\cdot$10$^6$ photons/pulse).}
	\label{fig:6}
\end{figure}

{\bf Scenario (A)}: we consider a coherent light beam impinging
 on a free-running detector with dead time $\Delta$t. The number of photons, $n$, hitting the detector within $\Delta$t follows a Poissonian distribution, $p(n\mid\mu)$, where $\mu$ is the average number of photons.
The probability, $p_{det}$, that the detector is ready and detects the photon of interest implies that no photons have been detected in the previous $\Delta$t seconds so
$$
p_{det}=\eta\sum_{n=0}^{\infty}p(n\mid\mu)(1-p_{det})^{n}
$$
where $\eta$ is the detection efficiency (in the non-saturated regime). Figure \ref{fig:6} shows $p_{det}$ with $\mu$ ranging from 0.1 and 5 and $\eta=0.45$. Due to the high count-rate and the dead time, $p_{det}$ for a free-running module decreases with the number of photons in the incoming beam, while it remains constant for a gated detector (assuming no noise photons arriving during the gate). 

\begin{figure}[!h]
	\centering
	\includegraphics[width=7 cm]{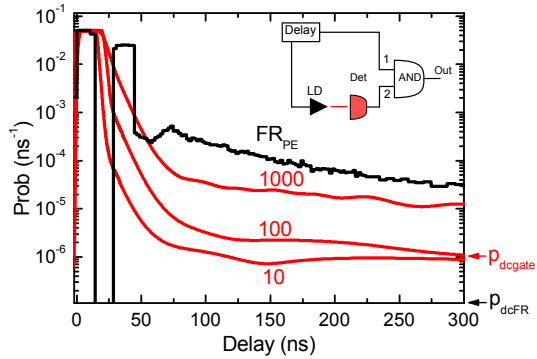}
	\caption{{\bf Scenario (B):} avalanche-triggering probability per ns vs the delay between the falling edge of the gate and the laser pulse (wavelength=808\,nm). In red, the gated module for different number of photons per pulse. In black, the free-running module (FR$_{PE}$) for 1000 photons per pulse. The red (black) arrow indicates the intrinsic dark count probability for the gated (FR$_{PE}$) module. {\bf Inset}: setup for time-resolved characterization.}
	\label{fig:7}
\end{figure}

{\bf Scenario (B)}: we compare a free-running Perkin Elmer SPCM-AQRH (FR$_{PE}$) with our gated module. These two modules are based on similar diodes. The arrival times of the detections in the free-running module are selected by an AND gate connected as shown in the inset of Fig.\ref{fig:7}. We use 808\,nm, 2\,ns pulses attenuated down to $10/100/1000$ photons per pulse. Note that we do not measure any significant variation in the decay time of the charge-persistence effect at different wavelengths.

Figure \ref{fig:7} shows the probability of having a click per ns as a function of the delay (defined as before). We see that the FR$_{PE}$, the black curve, goes to zero during the dead time. We also see a large peak due to the initially strong afterpulsing effect, another drop caused by the dead time, finally a slowly decreasing tail due to afterpulsing (note that the dead time is of about 32\,ns, however it appears shorter due to the width of the electrical outputs signal of 15\,ns).
The gated detector (red curve) shows a lower noise, in particular for lower pulse power. 

One could expect a smaller charge persistence effect for thin diode due to the smaller volume in the active region. Thus, we characterize a thin Si diode (CMOS, Id100 series by IDQuantique) in both free-running and gated mode of operations. This has also the advantage to compare the same diode in the two modes.

\begin{figure}[!h]
	\includegraphics[width=8 cm]{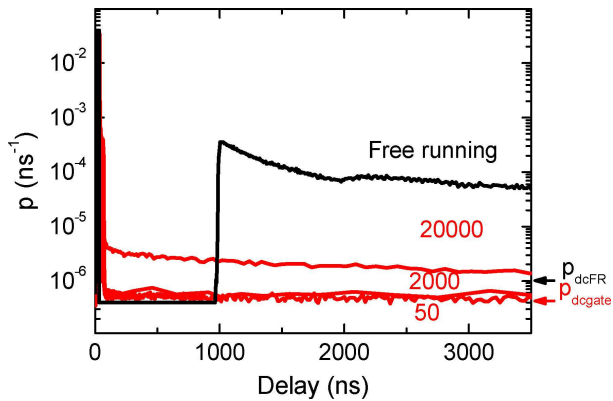}
	\caption{Time-resolved response (wavelength=655\,nm) for a thin diode by IDQuantique driven in free-running mode with 1000 photons per pulse and in gated mode with 20000/2000/50 photons per pulse. The arrows indicate the level of the intrinsic noise.}
	\label{fig:8}
\end{figure}

Figure \ref{fig:8} shows the time-resolved characterization of the diode driven in free-running mode (1000 photons per pulse) and in gated mode (50/2000/20000 photons per pulse). For technical reasons we cannot reduce the dead time in free-running mode below 1\,$\mu$s. The result confirms that gating has a direct advantage with respect to free-running operation in scenario {\bf (B)}. Indeed the thin diode is less affected by the charge-persistence noise (this effect can be seen only for pulses with more than 20000 photons per pulse). Similar results have been observed very recently for another thin diode\cite{DallaMoraAPL100}. We believe that the reduction of the absorption region is the main cause for this difference.\\
\section{Conclusions}
In conclusion, we report a Si-SPAD gated detector based on a commercially available diode. The detector achieves 45\%$\pm$5\% of efficiency at 808 nm with 2$\cdot$10$^{-6}$ dark count per ns at -30 V of excess bias and -30$^{\circ}$C. We showed the advantages of the gated mode for two measurement scenarios. We identified the charge persistence as the dominant source of excess noise and showed that this is less important for thin diodes.

\section*{Acknowledgements}
We thank R. Thew, I. Usmani, N. Walenta and N. Bruno for useful discussions and the Swiss NCCR Quantum Photonics (Technology transfer project) for financial support.

\end{document}